\documentclass[prl,amsfonts,amssymb,amsmath,showpacs]{revtex4}
\usepackage{graphicx}
\usepackage{bm}
\usepackage{ulem}

\begin{document}

\title{Excitonic Mott transition in double quantum wells}
\author{V.V.~Nikolaev}
\affiliation{School of Physics, University of
Exeter, Stocker Road, Exeter EX4 4QL, United Kingdom}
\author{M.E.~Portnoi}
\affiliation{School of Physics, University of
Exeter, Stocker Road, Exeter EX4 4QL, United Kingdom}

\date{August 28, 2003}

\begin{abstract}
We consider an electron-hole system in double quantum wells
theoretically. 
We demonstrate that there is a temperature interval over which 
an  abrupt  jump  in  the  value of the ionization  degree
occurs with an increase of the carrier density or temperature.
The opposite effect --- the collapse of the ionized electron-hole
plasma into an insulating exciton system --- should occur at 
lower densities.
In addition, we predict that under certain conditions there
will be a sharp decrease of the ionization degree with increasing
temperature --- the anomalous Mott transition. 
We discuss how these effects could be observed experimentally. 
\end{abstract}

\pacs{71.35.-y, 71.35.Ee, 73.21.Fg}

\maketitle

It was suggested by Mott \cite{Mott} that an 
increase of the temperature or density
of an insulating system of excitons may lead to 
avalanche ionization as a result of screening and 
k-space filling.
Despite significant efforts 
\cite{Balslev,ZimmermannMott,KochelapKuznetsov,LozovikBerman,DePalo}, 
there is still no
firm theoretical understanding of this effect, and the 
existence of the transition is disputed.

In bulk indirect-bandgap semiconductors, experimental
observation of the Mott transition is prevented 
by the formation of electron-hole liquid (EHL) droplets \cite{Rice}.
The absence of low-temperature phase 
transitions in direct semiconductors and single quantum wells
is believed to be caused by fast radiative recombination,
which does not allow enough time for the photo-generated electron-hole
plasma (EHP) to cool down.    
 
In recent years there has been increased attention 
towards double quantum wells (DQWs) --- heterostructures containing
two quantum wells (QWs) situated close enough so that the
Coulomb correlations between particles in different QWs
are significant. 
Electrons and holes from adjacent QWs form spatially 
indirect excitons (IXs) with very long lifetimes, due to
the small electron-hole (e-h) wavefunction overlap.
The experimentally confirmed 
long life-time and stability of the DQW indirect exciton 
\cite{Timofeev,Krivolapchuk, Butov, Butov2, Larionov}
suggests that the  formation of a medium-density cold exciton 
gas is possible. Under certain conditions this gas may undergo
a Mott transition.  

Another property beneficial for observation of the Mott transition 
in a DQW is that the spatially separated EHP
posses substantial electrostatic energy (Hartree term in the self-energy).
This leads to the fact that for large enough separation 
of quantum wells there is no minimum in the dependence of the
ground-state energy on the density, and thus 
the formation of e-h droplets is not 
possible \cite{LozovikBerman,ZhuLittlewood}.    

The purpose of this Letter is to investigate the 
possibility of a Mott transition in DQW systems and to encourage
further experimental research.
  
The main theoretical difficulty 
in investigation of the Mott transition is that it is 
inherently a medium-density effect, 
for which low-density and high-density approximations 
do not work. The correct approach would require 
careful handling of the Coulomb interaction depending on the
thermodynamic state of the system, which implies a 
self-consistent procedure.

Before studying the Mott transition, which is essentially 
a jump in the ionization degree of the EHP,
one should clarify the concept of  ionization degree
and exciton density at medium and high plasma densities. 
In the dilute case \cite{Portnoi:Ionization}, 
excitons can be considered as well-defined
Bose particles, and the problem of finding the ionization 
degree can be solved simply by counting electrons and holes in 
bound states pairwise. This approach, however,
runs into trouble when the density of carriers increases
and the exciton wavefunctions start to overlap.
Furthermore, this simple chemical picture is not applicable if the 
scattering states are occupied, since counting particles in scattering 
states pairwise leads to divergences.

Therefore, it appears more practical to formulate the theory 
in terms of primary quasi-particles, i.e. in terms of electrons
and holes. Instead of considering excitons in the system 
we investigate the single-particle density at given temperature 
\mbox{$T=1/\beta$} and quasi-Fermi energies $\xi_e$ and $\xi_h$.
To this end we use many-body theory for a system of 
interacting quasi-particles within the ladder approximation.
Starting with the self-energy in the quasi-particle
ladder approximation, which is based on the  
electron-electron (e-e) and e-h 
statically screened pair-interaction series, after of a lengthy
derivation \cite{Thesis} similar to that of Zimmermann and 
Stolz for the 3D case \cite{ZimmermannStolz:85}, we obtain an 
expression for the electron density in the following form: 
\begin{equation}
n_e=n_e^0(\xi_e)+n^{ee}(\xi_e)+n^{eh}(\xi_e+\xi_h),
\label{H1}
\end{equation}
where $n_e^0=m_e/(\beta\pi\hbar^2)\ln(1+\exp(\beta\xi_e))$ is
the density of free quasi-particles, and $n^{ee}$ ($n^{eh}$)
originates from the e-e (e-h) interaction:
\begin{eqnarray}
\nonumber\lefteqn{n^{ab}=\frac{2}{\beta\pi\hbar^2}
\sum_{m=-\infty}^{+\infty}
\left[(1-\delta_{ab})
\sum_n M_{eh}L_{eh}
\left(\epsilon_{mn}\right)\right.}\\
&&\left.+
\frac{1}{\pi}M_{ab}\lambda^{ab}_m\int_0^{\infty}dk
L_{ab}
\left(\frac{\hbar^2k^2}{2m_{ab}}
\right)2\sin^2\delta_m^{ab}(k)
\frac{d\delta_m^{ab}}{dk}
\right],
\label{Ultnab}
\end{eqnarray}
Here, $m_e$ ($m_h$) is the electron (hole) effective mass,
$M_{ab}=m_a+m_b$, $m_{ab}=m_am_b/(m_a+m_b)$, 
$\lambda^{ab}_m=1-\delta_{ab}(-1)^m/2$, and
$L_{ab}(\epsilon)=
-\ln\left[1-\exp\left(\beta({\xi_a+\xi_b-\epsilon})\right)
\right]$. The exciton levels $\epsilon_{mn}$
and generalized phase shifts $\delta_m^{ab}$ 
are obtained as the poles and the complex
arguments of the $m$th Fourier
components of the T-matrix.
The T-matrix equation
\begin{equation}
T^{ab}({\bf q}, {\bf q}_t, {\bf q}', \Omega_{ab})=
-V_s^{ab}({\bf q}'-{\bf q})-
\sum_{\bf q^*}
V_s^{ab}({\bf q}'-{\bf q})
\frac{1-f_a({\bf q}^*)-f_b({\bf q}_t-{\bf q}^*)}
{\epsilon_a({\bf q}^*)+\epsilon_b({\bf q}_t-{\bf q}^*)-\Omega_{ab}}
T^{ab}({\bf q}^*, {\bf q}_t, {\bf q}', \Omega_{ab})
\label{TmatrixEq}
\end{equation}
is solved using the factorized-potential 
approximation \cite{Thesis,ZimmermannStolz:85}.
Here, $\epsilon_{e(h)}$ are the electron (hole) quasi-particle
energies and 
$f_{e(h)}(\epsilon)=[\exp(\beta(\epsilon-\xi_{e(h)}))+1]^{-1}$
are Fermi distribution functions.
The major advantage of the mass action law
for quasi-two-dimensional (2D) systems, 
in the form of Eqs.~\eqref{H1}-\eqref{TmatrixEq}, over previously used
approaches \cite{Portnoi:Ionization} is that it is not confined to 
low densities  and accounts for k-space filling.
It is essential to include the effects of k-space filling
into a theory of the Mott transition since in two dimensions 
any weak symmetric potential has a bound state.

The next effect which should be carefully dealt with is screening. 
Using the Green's function technique, we have derived the following 
general expressions for the linearly-screened Coulomb interaction 
in quasi-2D systems \cite{paper02}:
\begin{equation}
V_s^{ab}=\eta_{a}\eta_{b}v(q)\left[F^{ab}(q)-v(q)\pi^{\bar{b}\bar{a}}
(q,\Omega)S(q)\right]
/D(q,\Omega).
\label{Vs}
\end{equation}
Here, $v=2\pi e^2/q$ is the bare Coulomb potential,
$F^{ab}$ are the form factors, $\pi^{ab}$ are the polarization
functions,
$\eta_{e(h)}=1(-1)$, $\bar{e}(\bar{h})=h(e)$, 
$S=F^{ee}F^{hh}-(F^{eh})^2$, 
$D=1-v\Pi_1+v^2\Pi_2S$, 
where $\Pi_1=\pi^{ee}F^{ee}+\pi^{hh}F^{hh}-2\pi^{eh}F^{eh}$,
and $\Pi_2=\pi^{hh}\pi^{ee}-(\pi^{eh})^2$.

Retaining only the ladder diagrams in the expressions of
polarization functions \cite{RopkeDer:A1}, and using an approximation 
similar to the plasmon-pole approximation for
the RPA polarization function $\pi^{RPA}$, the following expressions
are obtained: $\pi^{ee}=\pi^{RPA}+\pi^{ee}_{PP}+\pi^{eh}_{PP}$,
$\pi^{eh}=\pi^{eh}_{PP}$, where
\begin{equation}
\pi^{ab}_{PP}(q,\Omega)=-\frac{\hbar^2q^2}{M_{ab}}
\frac{n^{ab}}
{\frac{2\pi e^2\hbar^2}{\kappa M_{ab}}
\frac{n^{ab}}{q^{ab}_s}q^2+\frac{\hbar^4}
{4M^2_{ab}}q^4-\hbar^2\Omega^2}
\label{PIee001},
\end{equation} 
with the screening parameter 
\begin{equation}
q^{ab}_s=(1+\delta_{ab})\frac{2\pi e^2}{ \kappa}\frac{dn^{ab}}{d(\xi_a+\xi_b)}.
\label{qsab}
\end{equation}
Inspection of Eq.~\eqref{Vs} shows \cite{paper02} that in spatially 
separated systems screening by IXs is of the 
same order as screening by the free-carrier plasma, and this is due 
to the well-defined dipole momentum of the exciton.
This questions the validity of applying previously developed
theories of quasi-2D screening, which ignore screening by IXs,
to DQW systems. 

In the quasi-particle approximation, the chemical potential $\mu$ 
is given as the sum of the Fermi energy and the average quasi-particle shift
$\Delta_{e(h)}$. Eqs.~\eqref{H1}--\eqref{qsab} show that the density 
is a function of the Fermi energies only
(the relation between $\xi_e$ and $\xi_h$ is established by
the charge neutrality).
However, it is the quasi-particle shifts (sometimes called the
exchange-correlation part of the chemical potential) that can lead to
negative values of $d\mu/dn$, which in this case will mean
the formation of droplets. 
As was already mentioned, for large enough separations
the Hartree term of the self-energy prevents a first-order
e-h (or exciton) liquid phase transition; in this 
case $\mu$ is a monotonic function of the quasi-Fermi energy
and the shifts can be completely ignored \cite{Note}. 
This justifies the omission of
the exciton-exciton interaction
in the derivation of Eq.~\eqref{H1}, 
since it will mainly contribute to the 
positive quasi-particle shifts, which are due to dipole-dipole 
repulsion.

The e-h correlation term 
of the total density, $n^{eh}$, in Eqs.~\eqref{H1} and 
\eqref{Ultnab} depends on the sum of $\xi_e$ and 
$\xi_h$, and thus it is unchanged with a variation
of the local field. This represents insulating
behavior, whereas the first two terms in Eq.~\eqref{H1}
depend on $\xi_e$ alone and show metallic behavior. 
Therefore, it is reasonable to introduce
the ionization degree of the EHP as
$\alpha=1-n^{eh}/{n_e}=1-n^{eh}/{n_h}$, yielding
a mass action law which is not restricted 
to the low-density regime.

\begin{figure}
\begin{center}
\includegraphics[width=12cm,keepaspectratio]{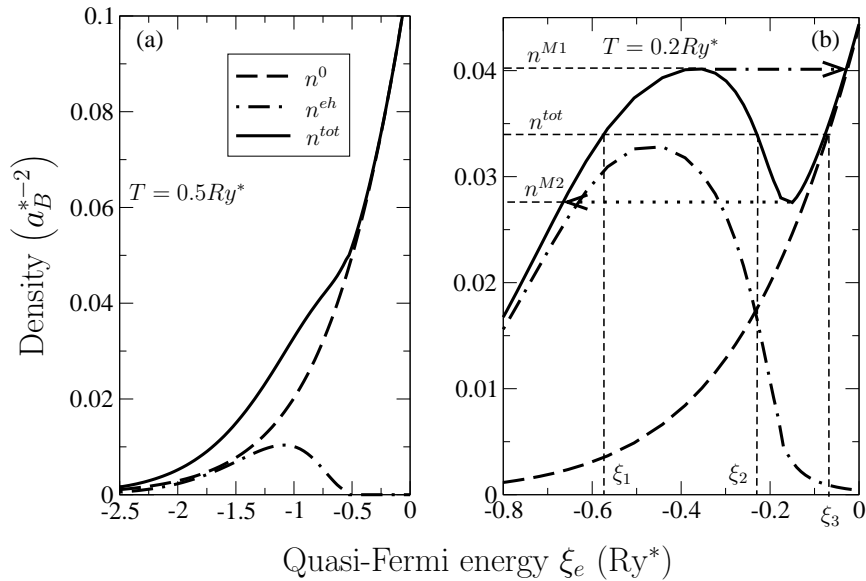}
\end{center}
\caption{Dependence of the different components of the carrier density
in a DQW on the electron quasi-Fermi energy.}
\label{fig1}
\end{figure}

\begin{figure}
\begin{center}
\includegraphics[width=10cm,keepaspectratio]{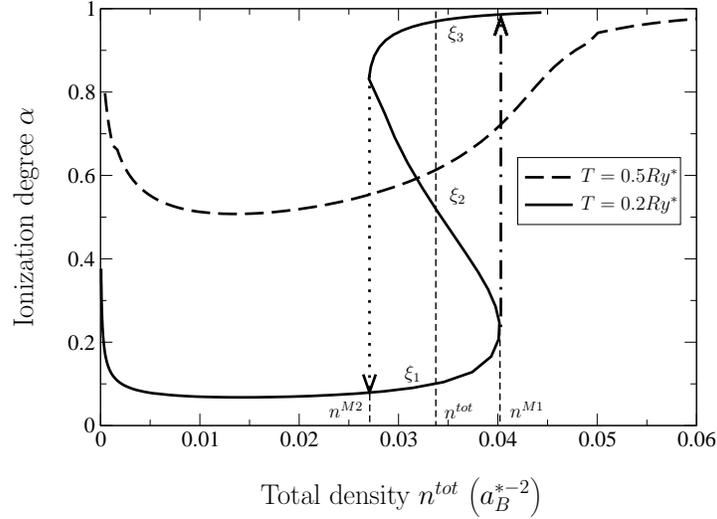}
\end{center}
\caption{Ionization degree for an EHP in a DQW at two different 
temperatures. Arrows show possible phase transitions.}
\label{fig2}
\end{figure}

We use a self-consistent procedure to calculate
the parameters of the spatially-separated EHP. 
Screening is considered statically, i.e.  
$\Omega=0$ is assumed in Eq.~\eqref{Vs}.
Fig.~\ref{fig1} shows the dependence of the free 
quasi-particle ($n^0$), e-h ($n^{eh}$) and total 
(\mbox{$n^{tot}=n^{0}+n^{ee}+n^{eh}$})
density in the DQW (the width of the QWs is 0.25 $a^*_B$ and the 
distance between them is 0.2 $a^*_B$) on the electron quasi-Fermi energy.
Here, $a^*_B$ and ${\rm Ry}^*$ are the Bohr radius and Rydberg of the bulk 
(three-dimensional) exciton.
The corresponding dependence of the ionization degree is 
plotted in Fig.~\ref{fig2}. 
One can see that for high temperature (0.5~${\rm Ry}^*$)
the total density increases monotonically,
whereas in the case of low temperature
(0.2~${\rm Ry}^*$) there is a region
of values of the Fermi energy in which a growing
free quasi-particle density is unable to
compensate for the decrease of the 
correlated e-h component, $n^{eh}$.
In this region the total density curve
acquires a negative slope, and a local 
maximum ($n^{M1}$) and minimum ($n^{M2}$) 
of the total density as a function of $\xi_e$ appear.
The ionization degree dependence takes a `hysteresis' 
form (Fig.~\ref{fig2}).

If the density of the e-h pairs in the system $n^{tot}$
lies between $n^{M1}$ and $n^{M2}$, there are
three corresponding states of the system characterized 
by three different Fermi energies, $\xi_1$, $\xi_2$
and $\xi_3$ (see Figs.~\ref{fig1}b and \ref{fig2}).
The thermodynamically stable state is $\xi_1$ since it 
has the smallest free energy (and ionization degree, see Fig.~\ref{fig2}).
The intermediate state $\xi_2$ is unstable 
due to a negative compressibility, $d\mu/dn=(d\xi_e/dn)(d\mu/d\xi_e)<0$.
The  state $\xi_3$ with the highest $\alpha$ is quasi-stable, with a local
minimum of free energy at the given density and chemical potential. 

As the density gradually
increases from zero, at $n^{tot}=n^{M1}$ there will be 
an abrupt transition to the state with higher $\xi$, shown in 
Figs.~\ref{fig1}(b) and \ref{fig2} by a thick dash-dot arrow.  
Physically this situation means the following.
Increasing the density of the carriers in the DQW system,
one reaches a critical value $n^{M1}$, at which a slight 
increase of the carrier concentration results 
in a large jump in the ionization degree.
Effectively, this means a transition from a system consisting 
mainly of excitons to an almost completely ionized state. 
Clearly, this is what the Mott transition (avalanche ionization) 
is meant to be. 
There will be no phase separation as this first-order transition 
takes place between phases with different chemical
potentials, i.e. equilibrium phase co-existence is impossible. 

However, there is another transition mechanism.
If the initial state of the system is a high-density 
ionized EHP and the density is slowly decreased, then the 
abrupt transition into an insulating state (ionized EHP collapse)
may happen not at the Mott density $n^{M1}$, but at a lower 
density $n^{M2}$ (the thick dotted arrow in Figs.~\ref{fig1} 
and \ref{fig2}). 
The reason for the difference between these two densities can
be explained qualitatively as follows. When the system is in the
insulating state, the screening due to excitons is comparatively 
weak, and accumulation of a substantial exciton density is 
required in order to trigger the avalanche ionization. 
After the avalanche ionization occurs, the screening in the system 
is mainly due to free carriers, i.e. it is much stronger, and 
one has to step back in density much further in order
for the system to collapse into excitons.   

\begin{figure}
\includegraphics[width=10cm,keepaspectratio]{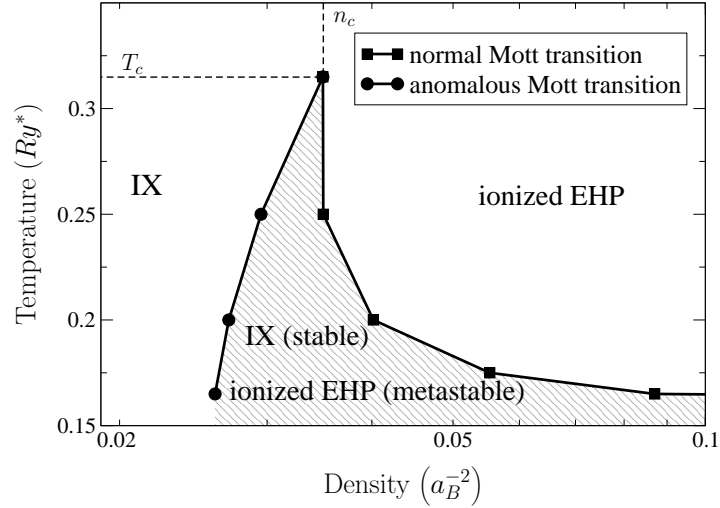}
\caption{The phase diagram for a DQW. The QW width is 
0.25~$a^*_B$, and the distance between QWs is 0.2~$a^*_B$.}
\label{fig3}
\end{figure}

In Fig.~\ref{fig3}, the phase diagram of a DQW structure is shown.
Both the critical temperature $T_c$ and density $n_c$ are 
experimentally accessible ($T_c$ is about 15 K and $n_c$ about
$2\times 10^{10}$ $\rm cm^{-2}$ for typical GaAs/AlGaAs structures). 
These parameters depend on the material and width of the 
individual QWs, as well as the interwell separation.
As the temperature is decreased, the avalanche ionization 
density increases drastically (we introduce 
the characteristic minimum temperature $T_{min}$ as the temperature
at which $n^{M1}=1.0~a^{*-2}_B$). 
The reason for this is that for lower temperatures 
the free-carrier density is reduced, and the weak screening by 
excitons and k-space filling requires higher densities to start 
the avalanche ionization process.

The phase diagram area below the transition curves
corresponds to the situation when the system can reside
in one of two states: a stable excitonic phase or metastable
ionized EHP. The excitation conditions, and the direction
and speed of the variation of the pumping intensity
and temperature influence the final state. 
It is interesting to consider the transition from a metastable 
ionized EHP to excitons (or back) by crossing the low-density 
transition curve with temperature variation. 
In this case, a sharp increase
of excitonic luminescence with a slow {\it increase} in temperature
or sharp disappearance of the excitonic line with a {\it decrease}
of temperature can take place. This direction of the transition 
is opposite to the normal Mott transition, and so we call it
the {\it anomalous} Mott transition.

\begin{figure}
\includegraphics[width=10cm]{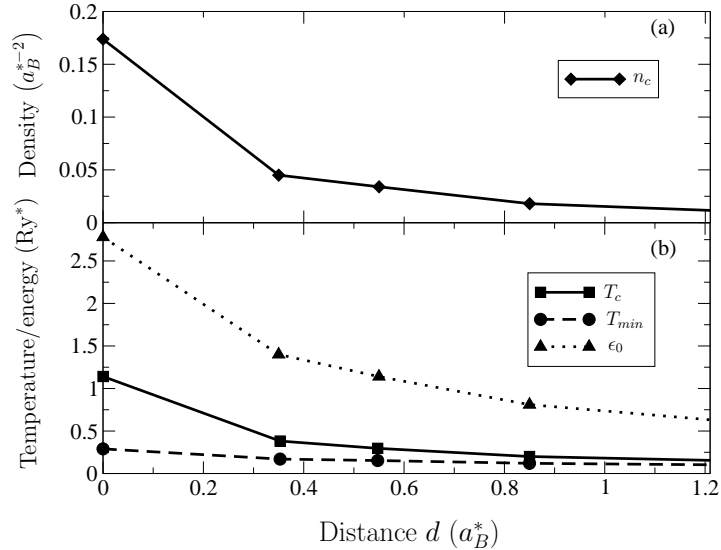}
\vspace{0.5cm}
\caption{The critical density (a) and temperature (b) as a 
function of e-h separation.
In (b) the binding energy of the IX and the 
characteristic minimum temperature are also shown.}
\label{fig4}
\end{figure}

In Fig.~\ref{fig4} the dependence of the critical parameters of 
the transition on the carrier separation is examined
(the distance $d$ is measured between the centers of the QWs,
and the points for a single-QW case, $d=0$, are shown for the 
sake of comparison.)
The critical density $n_c$ is decreasing with  
increasing e-h separation, which can be attributed 
to the increase of the IX radius. 
The critical temperature $T_c$ follows closely the 
decreasing trend of the IX binding energy $\epsilon_0$
(for large enough separation, $T_c\approx0.25\epsilon_0$).
The temperature $T_{min}$ is much less sensitive to the 
e-h separation. As a result, the temperature interval
in which avalanche ionization can occur shrinks with
increasing distance between QWs.

Encouraged by theoretical findings \cite{ZhuLittlewood,Lozovik75}, 
most experimental groups have been trying to obtain a quantum 
condensate of IXs \cite{Timofeev,Krivolapchuk,Butov,Butov2,Larionov}.  
In most cases this means concentrating the research effort on localized 
states and extremely low (below 1-3 K) temperatures.  
As we showed for delocalized excitons,
the observation of the normal Mott transition at such 
temperatures is difficult due to large critical density. 
At temperatures below $T_{min}$, the steady enhancement
of IX luminescence with an increase of pumping could be 
interpreted as a precursor to Bose condensation in some experiments.

Some experimental data \cite{Larionov}
show that localized states become filled at densities less than $3\times 10^9$
$\rm cm^{-2}$, in which case delocalized 
excitons start to dominate photoluminescence spectra. The same source
reports the observation of the insulator-metal
transition at densities less than $10^{11}$ $\rm cm^{-2}$. Unfortunately,
no information on the sharpness of the transition 
and the temperature dependence of critical density has been provided.

Timofeev {\it{et al.}} \cite{Timofeev} 
observed a sharp disappearance of the excitonic peak with a 
{\it decrease} of the temperature
at temperatures ${\rm{2K}}<T<{\rm{9K}}$ and low to moderate densities.
In our opinion, this may be interpreted as the anomalous Mott transition.
The measured transition curve is qualitatively
similar to the anomalous Mott transition curve in Fig.~\ref{fig3}.

We hope to encourage more experimental research 
which would lead to irrefutable observation of such
a fundamental effect as excitonic Mott transition.
Our theoretical results suggest the following method. 
The temperature should be set as low as possible and the pumping 
slowly increased to a high value, but so that the temperature 
kept constant.
Then, as the system is placed in the stable insulating state below 
the normal Mott transition curve in Fig.~\ref{fig2}, one should 
slowly increase the temperature. 
The Mott transition will manifest itself as the abrupt
elimination of the excitonic peak from the luminescence 
spectra or an increase of the longitudinal photo-conductance. 
Once the ionized state is achieved, one can try to 
observe the anomalous enhancement of exciton luminescence 
with a decrease of the density (photo-excitation).

We wish to thank I.~Galbraith for stimulating discussions
and D.G.W.~Parfitt for valuable advice.     

\normalem


\begin{thebibliography}{18}
\expandafter\ifx\csname natexlab\endcsname\relax\def\natexlab#1{#1}\fi
\expandafter\ifx\csname bibnamefont\endcsname\relax
  \def\bibnamefont#1{#1}\fi
\expandafter\ifx\csname bibfnamefont\endcsname\relax
  \def\bibfnamefont#1{#1}\fi
\expandafter\ifx\csname citenamefont\endcsname\relax
  \def\citenamefont#1{#1}\fi
\expandafter\ifx\csname url\endcsname\relax
  \def\url#1{\texttt{#1}}\fi
\expandafter\ifx\csname urlprefix\endcsname\relax\def\urlprefix{URL }\fi
\providecommand{\bibinfo}[2]{#2}
\providecommand{\eprint}[2][]{\url{#2}}

\bibitem[{\citenamefont{Mott}(1961)}]{Mott}
\bibinfo{author}{\bibfnamefont{N.~F.} \bibnamefont{Mott}},
  \bibinfo{journal}{Philos. Mag.} \textbf{\bibinfo{volume}{6}},
  \bibinfo{pages}{287} (\bibinfo{year}{1961}).

\bibitem[{\citenamefont{Balslev}(1978)}]{Balslev}
\bibinfo{author}{\bibfnamefont{I.}~\bibnamefont{Balslev}},
  \bibinfo{journal}{Solid State Commun.} \textbf{\bibinfo{volume}{27}},
  \bibinfo{pages}{545} (\bibinfo{year}{1978}).

\bibitem[{\citenamefont{Zimmermann}(1988)}]{ZimmermannMott}
\bibinfo{author}{\bibfnamefont{R.}~\bibnamefont{Zimmermann}},
  \bibinfo{journal}{Phys. Status Solidi B} \textbf{\bibinfo{volume}{146}},
  \bibinfo{pages}{371} (\bibinfo{year}{1988}).

\bibitem[{\citenamefont{Kochelap and Kuznetsov}(1990)}]{KochelapKuznetsov}
\bibinfo{author}{\bibfnamefont{V.~A.} \bibnamefont{Kochelap}} \bibnamefont{and}
  \bibinfo{author}{\bibfnamefont{A.~V.} \bibnamefont{Kuznetsov}},
  \bibinfo{journal}{Phys. Rev. B} \textbf{\bibinfo{volume}{42}},
  \bibinfo{pages}{7497} (\bibinfo{year}{1990}).

\bibitem[{Loz()}]{LozovikBerman}
\bibinfo{note}{Yu. E. Lozovik and O. L. Berman, Pis'ma Zh. Eksp. Teor. Fiz.
  {\bf 64}, 526 (1996) [JETP Lett. {\bf 64}, 573 (1996)]}.

\bibitem[{\citenamefont{De~Palo et~al.}(2002)\citenamefont{De~Palo,
  Moskalenko, and Zhmodikov}}]{DePalo}
\bibinfo{author}{\bibfnamefont{S.} \bibnamefont{De~Palo}},
  \bibinfo{author}{\bibfnamefont{F.} \bibnamefont{Rapisarda}},
  \bibnamefont{and} \bibinfo{author}{\bibfnamefont{G.}
  \bibnamefont{Senatore}}, \bibinfo{journal}{Phys. Rev. Lett.}
  \textbf{\bibinfo{volume}{88}}, \bibinfo{pages}{206401}
  (\bibinfo{year}{2002}).

\bibitem[{\citenamefont{Rice}(1977)}]{Rice}
\bibinfo{author}{\bibfnamefont{T.~M.} \bibnamefont{Rice}},
  \emph{\bibinfo{title}{{\rm in} Solid State Physics, {\rm edited by
  H.~Enrenreich, F.~Seitz, and D.~Turnbull}}} (\bibinfo{publisher}{Academic
  Press}, \bibinfo{address}{New York}, \bibinfo{year}{1977}),
  vol.~\bibinfo{volume}{32}, p.~\bibinfo{pages}{1}.

\bibitem[{\citenamefont{Timofeev et~al.}(2000)\citenamefont{Timofeev, 
Larionov, Grassi-Alessi
  Capizzi, and Hvam}}]{Timofeev}
\bibinfo{author}{\bibfnamefont{V.~B.}~\bibnamefont{Timofeev}},
  \bibinfo{author}{\bibfnamefont{A.~V.} \bibnamefont{Larionov}},
  \bibinfo{author}{\bibfnamefont{M.}~\bibnamefont{Grassi-Alessi}},
  \bibinfo{author}{\bibfnamefont{M.}~\bibnamefont{Capizzi}},
  \bibnamefont{and} \bibinfo{author}{\bibfnamefont{J.~M.}~\bibnamefont{Hvam}},
  \bibinfo{journal}{Phys. Rev. B} \textbf{\bibinfo{volume}{61}},
  \bibinfo{pages}{8420} (\bibinfo{year}{2000}).

\bibitem[{\citenamefont{Krivolapchuk et~al.}(2002)\citenamefont{Krivolapchuk,
  Moskalenko, and Zhmodikov}}]{Krivolapchuk}
\bibinfo{author}{\bibfnamefont{V.~V.} \bibnamefont{Krivolapchuk}},
  \bibinfo{author}{\bibfnamefont{E.~S.} \bibnamefont{Moskalenko}},
  \bibnamefont{and} \bibinfo{author}{\bibfnamefont{A.~L.}
  \bibnamefont{Zhmodikov}}, \bibinfo{journal}{Phys. Rev. B}
  \textbf{\bibinfo{volume}{64}}, \bibinfo{pages}{045313}
  (\bibinfo{year}{2002}).

\bibitem[{\citenamefont{Butov et~al.}(2002)\citenamefont{Butov, Lai, Ivanov,
  Gossard, and Chemla}}]{Butov}
\bibinfo{author}{\bibfnamefont{L.~V.} \bibnamefont{Butov}},
  \bibinfo{author}{\bibfnamefont{C.~W.} \bibnamefont{Lai}},
  \bibinfo{author}{\bibfnamefont{A.~L.} \bibnamefont{Ivanov}},
  \bibinfo{author}{\bibfnamefont{A.~C.} \bibnamefont{Gossard}},
  \bibnamefont{and} \bibinfo{author}{\bibfnamefont{D.~S.}
  \bibnamefont{Chemla}}, \bibinfo{journal}{Nature}
  \textbf{\bibinfo{volume}{417}}, \bibinfo{pages}{47} (\bibinfo{year}{2002}).

\bibitem[{But()}]{Butov2}
\bibinfo{note}{L. V. Butov, A. C. Gossard, and D. S. Chemla, Nature {\bf 418},
  751 (2002); D. Snoke {\it et al.}, Nature {\bf 418}, 754 (2002)}.

\bibitem[{Lar()}]{Larionov}
\bibinfo{note}{A. V. Larionov {\it et al.}, Pis'ma Zh. Eksp. Teor. Fiz. {\bf
  75}, 689 (2002) [JETP Lett. {\bf 75}, 570 (2002)]; A.~A. Dremin {\it et al.},
  Pis'ma Zh. Eksp. Teor. Fiz. {\bf 76}, 526 (2002) [JETP Lett. {\bf 76}, 450
  (2002)]}.

\bibitem[{\citenamefont{Zhu et~al.}(1995)\citenamefont{Zhu, Littlewood,
  Hybertsen, and Rice}}]{ZhuLittlewood}
\bibinfo{author}{\bibfnamefont{X.}~\bibnamefont{Zhu}},
  \bibinfo{author}{\bibfnamefont{P.~B.} \bibnamefont{Littlewood}},
  \bibinfo{author}{\bibfnamefont{M.}~\bibnamefont{Hybertsen}},
  \bibnamefont{and} \bibinfo{author}{\bibfnamefont{T.}~\bibnamefont{Rice}},
  \bibinfo{journal}{Phys. Rev. Lett.} \textbf{\bibinfo{volume}{74}},
  \bibinfo{pages}{1633} (\bibinfo{year}{1995}).

\bibitem[{\citenamefont{Portnoi and Galbraith}(1999)}]{Portnoi:Ionization}
\bibinfo{author}{\bibfnamefont{M.~E.} \bibnamefont{Portnoi}} \bibnamefont{and}
  \bibinfo{author}{\bibfnamefont{I.}~\bibnamefont{Galbraith}},
  \bibinfo{journal}{Phys. Rev. B} \textbf{\bibinfo{volume}{60}},
  \bibinfo{pages}{5570} (\bibinfo{year}{1999}).

\bibitem[{\citenamefont{Nikolaev}(2002)}]{Thesis}
\bibinfo{author}{\bibfnamefont{V.~V.} \bibnamefont{Nikolaev}}, Ph.D. thesis,
  \bibinfo{school}{University of Exeter} (\bibinfo{year}{2002}).

\bibitem[{\citenamefont{Zimmermann and Stolz}(1985)}]{ZimmermannStolz:85}
\bibinfo{author}{\bibfnamefont{R.}~\bibnamefont{Zimmermann}} \bibnamefont{and}
  \bibinfo{author}{\bibfnamefont{H.}~\bibnamefont{Stolz}},
  \bibinfo{journal}{Phys. Status Solidi B} \textbf{\bibinfo{volume}{131}},
  \bibinfo{pages}{151} (\bibinfo{year}{1985}).

\bibitem[{\citenamefont{Nikolaev et~al.}(2003)\citenamefont{Nikolaev, Avrutin,
  and Portnoi}}]{paper02}
\bibinfo{author}{\bibfnamefont{V.~V.} \bibnamefont{Nikolaev}},
  \bibinfo{author}{\bibfnamefont{E.~A.} \bibnamefont{Avrutin}},
  \bibnamefont{and} \bibinfo{author}{\bibfnamefont{M.~E.}
  \bibnamefont{Portnoi}} (\bibinfo{year}{2003}), \bibinfo{note}{unpublished}.

\bibitem[{\citenamefont{R{\"{o}}pke and Der}(1979)}]{RopkeDer:A1}
\bibinfo{author}{\bibfnamefont{G.}~\bibnamefont{R{\"{o}}pke}} \bibnamefont{and}
  \bibinfo{author}{\bibfnamefont{R.}~\bibnamefont{Der}},
  \bibinfo{journal}{Phys. Status Solidi B} \textbf{\bibinfo{volume}{92}},
  \bibinfo{pages}{501} (\bibinfo{year}{1979}).

\bibitem[{Not({\natexlab{a}})}]{Note}
\bibinfo{note}{Rigorous investigation of the transition from the EHL regime to
 the homogeneous EHP regime with increasing electron-hole separation is beyond 
 the scope  of this paper, but it should be noted that when the EHL transition 
 is not prohibited by the large Hartree energy (i.e. in single or very closely
 situated QWs), it may not be achievable due to fast radiative recombination.}

\bibitem[{Loz2()}]{Lozovik75}
\bibinfo{note}{Yu. E. Lozovik and V. I. Yudson, Pis'ma Zh. Eksp. Teor. Fiz.
  {\bf 22}, 556 (1975) [JETP Lett. {\bf 22}, 274 (1975)]}.

\end{thebibliography}
\end{document}